\documentclass[twocolumn,showpacs,preprintnumbers,amsmath,amssymb]{revtex4}
\usepackage{epsfig}
\usepackage{graphicx}
\usepackage{dcolumn}
\usepackage{bm}
\begin{document}
\title{Supersymmetry representation of Bose-Einstein condensation
of fermion pairs}%
\author{Alexander Olemskoi}
\email{alex@ufn.ru}
\author{Irina Shuda}
\email{shudaira@mail.ru}
 \affiliation{Institute of Applied Physics, Nat. Acad. Sci. of Ukraine, 58,
Petropavlovskaya St., 40030 Sumy, Ukraine\\ Sumy State University, 2,
Rimskii-Korsakov St., 40007 Sumy, Ukraine}

\date{ \today }

\begin{abstract}
We consider supersymmetry field theory with supercomponents being the square
root of the Bose condensate density, the amplitude of its fluctuations and
Grassmannian fields related to the Fermi particles density. The fermion number
is demonstrated to be conserved in degenerated Fermi-Bose mixtures with
unbroken supersymmetry when the system is invariant with respect to inversion
of the time arrow. We show the supersymmetry breaking allows one to derive
field equations describing behavior of real Bose-Fermi mixtures. Solution of
related field equations reveals the cooling of homogeneously distributed
fermions gives first spontaneous rise to strong inhomogeneous fluctuations,
while the Bose condensate appears at a lower temperature dependent of the
fermion density.
\end{abstract}


\maketitle

\section{Introduction}\label{Sec.1}

Supersymmetry is one of the most beautiful and productive conception of
contemporary physics which has been proposed for description of microworld (see
Ref. \cite{Cooper} to review supersymmetry applications to the quantum
mechanics, the survey \cite{Ef} deals with disorder metals, and book \cite{GSW}
relates to the theory of superstrings). Basing on idea proposed in the work
\cite{23} and developed in Refs. \cite{12} and \cite{016}, it has been shown
that the supersymmetry can be used effectively to present fluctuative fields
determining a picture of phase transition with symmetry breaking in
non-equilibrium condensed matter \cite{2_13a} and random heteropolymers
\cite{2_13b}. Along this line, the Martin-Siggia-Rose method \cite{30} has been
used as a basis permitting to combine stochastic fields into supersymmetry
construction \cite{Zinn}. One can perceive the supersymmetry approach is
developed along two main lines: the former is based on the quantum operator
representation \cite{Cooper,Ef}, while the latter considers evolution of
supersymmetric field \cite{GSW,23,12,016,2_13a,2_13b,30,Zinn}.

Initially, the supersymmetry had been proposed as a symmetry that relates
bosons and fermions in elementary particle physics \cite{GoLi1,VoAk,WZ}. Along
this line, the supersymmetric string theory is a unique theory expected to give
a unified description of all interactions in nature \cite{GSW}, however none of
the superpartners of any known elementary particles has been found in
experiments so far. Therefore, it is very important to study the supersymmetry
breaking. Such an opportunity is opened with experimental progress in atomic
mixtures of ultracold Bose and Fermi atoms \cite{Pit}. Theoretically, an
ultracold superstring model was constructed \cite{6}, as well as an exactly
soluable model of one-dimensional Bose-Fermi mixture was investigated \cite{7}.
According to Ref. \cite{8}, the supersymmetry is always broken either
spontaneously or by a chemical potential difference between bosons and
fermions. This article is devoted to consideration of the Bose-Fermi mixture
within field-theoretical supersymmetry approach \cite{Zinn}.

The outline of the paper is as follows. In section \ref{Sec.2}, we adduce main
field-theoretical statements based on the generating functional method
introduced by Martin, Siggia and Rose. Making use of this method allows us to
write down both supersymmetric Lagrangian of the problem and Euler equations
related. Section \ref{Sec.3} is devoted to derivation of the field equations
for the superfield components being the most probable values of the square root
of the Bose condensate density, the amplitude of its fluctuations and the most
probable Grassmannian fields giving the Fermi particles density. Combining
above equations, we show the fermion number is conserved in supersymmetrically
degenerated systems whose superspace is invariant with respect to a rotation.
According to section \ref{Sec.4}, the supersymmetry invariance is broken by
means of fixation of such rotation that results in loss of the Grassmannian
components invariance with respect to the time inversion due to the
Bose-Einstein condensation. We show the supersymmetry breaking allows one to
derive field equations describing behavior of real Bose-Fermi mixtures. Section
\ref{Sec.5} is devoted to discussion of obtained results and Appendix contains
details of calculations at derivation of the field equations.

\section{Main field-theoretical statements}\label{Sec.2}

We consider attractive Fermi system characterised by the pair of conjugate wave
functions $\psi({\bf r},t)$, ${\overline\psi}({\bf r},t)$ and Bose condensate
with the density $n({\bf r},t)$ where $\bf r$, $t$ are coordinate and time,
respectively. Behavior of the Bose subsystem is presented by the fluctuating
order parameter
\begin{equation}
x({\bf r},t):=\sqrt{n({\bf r},t)}{\rm e}^{{\rm i}\phi({\bf r},t)} \label{2}
\end{equation}
with $\phi({\bf r},t)$ being a condensate phase. Within framework of the
standard field-theoretical scheme \cite{Zinn}, the system evolution is
described by the Langevin equation
\begin{equation}
\dot{x}({\bf r},t) - D\nabla^2 x = -\gamma~ {\partial F\over\partial x} +
\zeta({\bf r},t). \label{3}
\end{equation}
Here, dot stands for the time derivative, $\nabla\equiv\partial/\partial {\bf
r}$, $D$ and $\gamma$ are diffusion and kinetic coefficients, respectively,
$F(x)$ is specific free energy of condensate, $\zeta({\bf r}, t)$ is stochastic
addition defined by the white noise conditions
\begin{equation}
\left<\zeta({\bf r},t)\right>=0, \qquad \left<\zeta({\bf r},t)\zeta({\bf
0},0)\right>=\gamma T\delta({\bf r})\delta(t) \label{4}
\end{equation}
where angle brackets notice averaging over system states scattered with
variance $T$ being the temperature measured in energy units. With introduction
of the scales $t_s\equiv (\gamma T)^2/D^3$, $r_s\equiv \gamma T/D$, $F_s\equiv
D^3/\gamma^3T^2$, $\zeta_s\equiv D^3/(\gamma T)^2$ for time $t$, coordinate
${\bf r}$, specific free energy $F$ and stochastic force $\zeta$, respectively,
the equation of motion (\ref{3}) takes the simplest form
\begin{equation}
\dot{x}({\bf r},t)=-{\delta\mathcal{F}\over\delta x}+\zeta({\bf r},t)
\label{2.3}
\end{equation}
where short denotion of the variotinal derivative
\begin{equation}
{\delta\mathcal{F}\over\delta x} \equiv {\delta\mathcal{F}\{ x({\bf r},t) \}
\over\delta x({\bf r},t)} ={\partial F(x)\over\partial x} -\nabla^2 x
 \label{2.4}
\end{equation}
is used for the Ginzburg-Landau model
\begin{equation}
\mathcal{F}\{x\}\equiv\int\left[F(x) +{1\over 2} (\nabla x)^2\right]{\rm d}{\bf
r}. \label{2.4a}
\end{equation}

Along the standard line \cite {Zinn}, our approach is stated on the generating
functional
\begin{equation}
Z\{u({\bf r},t)\}=\int \!\!Z\{x\}\exp\left(\int\! u x \, {\rm d}{\bf r}\, {\rm
d}t\right) {\rm D}x \label{2.6}
\end{equation}
being determined by the partition functional
\begin{equation}
Z\{x({\bf r},t)\}:=\left<\prod_{({\bf r},t)}\delta\left\{\dot{x}+{\delta F\over
\delta x}-\zeta\right\}{\rm det}\left|{\delta\zeta\over\delta x}\right|\right>
\label{2.7}
\end{equation}
where argument of the $\delta$-functional takes into account the equation of
motion (\ref{2.3}) and determinant is the Jacobian of the transition from the
noise field $\zeta({\bf r},t)$ to the order parameter $x({\bf r},t)$ over whose
distribution the continuous integration in the definition (\ref{2.6}) is
carried out.

Within the simplest case of the \^Ito calculus, the Jacobian determinant equals
one and the expression (\ref{2.7}) arrives at the pair of the Bose field only
\cite {2_13b}. Much more interesting situation is generated by the Stratonovich
calculus when the Jacobian
\begin{equation}
\det\left|\frac{\delta\zeta}{\delta x}\right|=\int \exp\left(\overline\psi
\frac{\delta\zeta}{\delta x} \psi\right){\rm d}^2 \psi, \quad {\rm d}^2\psi =
{\rm d} \psi~{\rm d}\overline \psi \label{2.33}
\end{equation}
is presented by Grassmannian conjugate fields $\psi({\bf r}, t)$ and
$\overline\psi ({\bf r}, t)$ which subject to the following conditions:
\begin{equation} \label{2.12a}
\begin{split}
\overline\psi \psi+\psi\overline\psi =0,\quad \int {\rm d}\psi=0;\\
\int\psi~{\rm d}\psi=1,\quad \int {\rm d}\overline\psi=0, \quad
\int\overline\psi~{\rm d}\overline\psi=1.
\end{split}
\end{equation}
Then, after generalized Laplace transform of the $\delta$-functional in
Eq.(\ref{2.7}) we obtain the supersymmetry Lagrangian
\begin{equation}
\begin{split}
{\cal L}(x,p,\psi,\overline\psi)= \left(p\dot x - {p^2\over 2} + {\delta F\over
\delta x}p\right)\\ - \overline\psi \left( {\partial \over \partial
t}+{\delta^2 F \over \delta x^2} \right)\psi \label{2.34}
\end{split}
\end{equation}
with a ghost field $p({\bf r},t)$. Introducing the four-component superfield
\begin{equation} \Phi:= x +\overline\theta\psi
+\overline\psi\theta + \overline\theta\theta p, \label{2.35}
\end{equation}
easily to convince that expression (\ref{2.34}) can be written in the canonical
supersymmetric form
\begin{equation}\label{2.36}
\begin{split}
{\cal L} =\int \Lambda {\rm d}^2 \theta,\quad {\rm d}^2 \theta \equiv {\rm d}
\theta{\rm d}\overline\theta,\\ \Lambda (\Phi) \equiv {1\over 2}(\overline
{\cal D} \Phi)\left( {\cal D}\Phi\right) + F(\Phi)
\end{split}
\end{equation}
where $\theta$, $\overline\theta$ are Grassmannian conjugate coordinates
defined by the properties
\begin{equation} \label{2.12b}
\begin{split}
\overline\theta\theta+\theta\overline\theta=0,\quad \int {\rm d}\theta=0;\\
\quad \int\theta{\rm d}\theta=1,\quad \int {\rm d}\overline\theta=0, \quad
\int\overline\theta{\rm d}\overline\theta=1
\end{split}
\end{equation}
being similar to Eqs.(\ref{2.12a}). Supersymmetry generators in Eq.(\ref{2.36})
are as follows:
\begin{equation}
{\cal D}:={\partial \over \partial \overline \theta}-2\theta
{\partial\over\partial t}, \qquad \overline {\cal D}:={\partial \over \partial
\theta}. \label{2.37}
\end{equation}

\section{Supersymmetry field equations}\label{Sec.3}

Variation of the action related to the supersymmetric Lagrangian (\ref{2.36})
over the superfield (\ref{2.35}) derives the supersymmetry Euler equation
\begin{equation}
{1\over 2} [\overline {\cal D},{\cal D}]\Phi={\delta\mathcal{F}\over \delta
\Phi} \label{2.38}
\end{equation}
where the square brackets notice commutation. As is shown in Appendix,
projection of Eq.(\ref{2.38}) onto the superspace axes $1$, $\overline\theta$,
$\theta$ and $\overline\theta\theta$ arrives at the explicit form of the
equations of motion:
\begin{equation}
\dot\eta -\nabla^2\eta=- {\partial F\over \partial\eta} + \varphi,
\label{2.39a}
\end{equation}
\begin{equation}
\dot\varphi +\nabla^2\varphi = {\partial^2 F\over
\partial\eta^2}~\varphi - {\partial^3 F\over\partial\eta^3}~
\overline\Psi\Psi, \label{2.39b}
\end{equation}
\begin{equation}
\dot\Psi -\nabla^2\Psi =- {\partial^2 F\over\partial\eta^2}~\Psi, \label{2.39c}
\end{equation}
\begin{equation}
-\dot{\overline\Psi}-\nabla^2\overline\Psi=- {\partial^2
F\over\partial\eta^2}~\overline\Psi. \label{2.39d}
\end{equation}
Since minimal action relates to the most probable realizations of the
superfield (\ref{2.35}), solutions of equations (\ref{2.39a}) -- (\ref{2.39d})
determine the most probable components $x^{(max)}\equiv\eta$ \footnote{For
simplicity, we suppose the most probable value $\eta=\sqrt{n}$ of the order
parameter (\ref{2}) to be totally real.}, $p^{(max)}\equiv\varphi$,
$\psi^{(max)}\equiv\Psi$ and $\overline\psi^{(max)}\equiv\overline\Psi$. The
first of these equations takes the form of the Langevin equation (\ref{2.3})
that shows a ghost field $\varphi\equiv p^{(max)}$ represents the most probable
realization of the fluctuation amplitude $\zeta$. Specific peculiarity of the
field $\varphi$ is that gradient term in the governing equation (\ref{2.39b})
has inverse sign, so that inhomogeneity in the space distribution of the most
probable fluctuation increases in the course of the time until non-linearity
stabilizes its amplitude.

Another feature consists in the Grassmannian conjugation of the equations
(\ref{2.39c}) and (\ref{2.39d}) which coincide if the time arrow is inverted in
one of them. Thus, one can conclude that the pair of conjugate fields $\Psi$
and $\overline\Psi$ describes evolution of the Fermi particle and antiparticle
for which the time runs in opposite directions. Combining Eqs. (\ref{2.39c})
and (\ref{2.39d}) arrives at the continuity equation
\begin{equation}
\dot\rho+\nabla{\bf j}=0
 \label{2.40}
\end{equation}
for the fermion density
\begin{equation}
\rho:=\overline\Psi\Psi
 \label{2.40a}
\end{equation}
and the current related
\begin{equation}
{\bf j}:=\nabla\overline\Psi\Psi-\overline\Psi\nabla\Psi.
 \label{2.40b}
\end{equation}
The equation (\ref{2.40}) expresses the conservation law of the Fermi particles
number in supersymmetric system whose state space spanned onto axes $1$,
$\overline\theta$, $\theta$ and $\overline\theta\theta$ is invariant with
respect to a direction choice.

\section{Supersymmetry breaking}\label{Sec.4}

To break above invariance we shall follow to the Bogolyubov method of
quasi-averages, according to which taking off a system degeneration is provided
by switching an infinitesimal source type of slight magnetic field in magnets
\cite{Bog}. In our case, the role of such a field is played by the conjugate
Grassmannian components $\overline\theta\Psi$ and $\overline\Psi\theta$ the
superfield (\ref{2.35}) which relate to forward and backward directions of the
time arrow. Formally, we should replace the superfield (\ref{2.35}) by the
transformed field
\begin{equation}
\widetilde\Phi:={\rm e}^{-\overline\theta\Psi}\Phi{\rm
e}^{\overline\theta\Psi}. \label{2.50}
\end{equation}
Writing this superfield in the explicit form
\begin{equation} \label{2.50a}
\widetilde\Phi=x+\left(1-x\right)\overline\theta\Psi
+\left(1+x\right)\overline\Psi\theta
+\left(p+x\rho\right)\overline\theta\theta,
\end{equation}
easily to see that transformation (\ref{2.50}) squeezes the axis
$\overline\theta\Psi$ and stretches the axis $\overline\Psi\theta$ by the same
value $x$ being the order parameter of the Bose condensate, while the axis
$\overline\theta\theta$ is stretched by the value $x\rho$ proportional to both
order parameter and density of the Fermi particles $\rho=\overline\Psi\Psi$. In
any case, above transformation breaks the supersymmetry, so that the Euler
superequation (\ref{2.38}) is reduced to the following components (see
Appendix):
\begin{equation}
\frac{\partial\eta}{\partial t}-\nabla^2\eta=-{\partial F\over
\partial\eta} + \varphi+\eta\rho, \label{2.51a} \end{equation}
\begin{equation} \label{2.51b}
\begin{split}
&\frac{\partial}{\partial t}\left(\varphi+\eta\rho\right)+\nabla^2\varphi\\&=
{\partial^2 F\over
\partial\eta^2}\left(\varphi+\eta\rho\right)-{\partial^3 F\over \partial\eta^3}
\left(1-\eta^2\right)\rho,\end{split}
 \end{equation}
\begin{equation}
\Psi\frac{\partial}{\partial
t}\ln\left[\left(1-\eta\right)\Psi\right]-\nabla^2\Psi = - {\partial^2 F \over
\partial\eta^2}\Psi, \label{2.51c}
\end{equation}
\begin{equation}
-\overline\Psi\frac{\partial}{\partial
t}\ln\left[\left(1+\eta\right)\overline{\Psi}\right]-\nabla^2\overline\Psi = -
{\partial^2 F \over \partial\eta^2}\overline\Psi. \label{2.51d}
\end{equation}
In contrast to Eqs. (\ref{2.39c}) and (\ref{2.39d}), the pair of equations
(\ref{2.51c}) and (\ref{2.51d}) becomes non-invariant with respect to the time
inversion due to the Bose-Einstein condensate appearance $(\eta\ne 0)$.
Combining the equations (\ref{2.51c}) and (\ref{2.51d}), one obtains
\begin{equation}
\rho\frac{\partial}{\partial
t}\ln\left[\left(1-\eta^2\right)\rho\right]+\nabla{\bf j}=0. \label{2.52}
\end{equation}
At steady-state condensation $({\bf j}={\rm const})$, Eq.(\ref{2.52}) arrives
at the relation
\begin{equation}
n=1-\frac{\rho_c}{\rho}
 \label{2.52a}
\end{equation}
where integration constant $\rho_c$ plays the role of a critical density of
fermions and one takes into account the definition (\ref{2}) according to which
$\eta^2=n$. The dependence (\ref{2.52a}) means the density $n$ of the Bose
condensate increases steadily from $n=0$ to $n=1$ with growth of the density
$\rho$ of Fermi particles above a critical value $\rho_c$.

\section{Discussion}\label{Sec.5}

Characteristic feature of our consideration consists in making use of the
supersymmetry field theory that is based on principle of the minimal
superaction $$S\{\Phi({\bf r},t)\}:=\int\mathcal{L}[\Phi({\bf r},t)]{\rm d}{\bf
r}{\rm d}t$$ whose values relate to maximal probability $$P\{\Phi({\bf
r},t)\}\propto\exp\left(-S\{\Phi({\bf r},t)\}\right)$$ in the system
distribution over the superfields (\ref{2.35}). As a result, the governing
equations (\ref{2.51a}) -- (\ref{2.51d}) determine the most probable Bose
components $\eta$, $\varphi$ and Fermi ones $\Psi$, $\overline\Psi$. Such a
description is differed crucially from the standard picture, within whose
framework observable values are determined in terms of averages over sets of
quantum states. It is worthwhile to stress though the most probable fields are
determined without averaging over quantum fluctuations, however they take into
account scattering over statistical states (see related averaging in
Eqs.(\ref{4})).

According to Eqs. (\ref{2.39c}) and (\ref{2.39d}), Fermi-Bose mixtures with
unbroken supersymmetry are invariant with respect to inversion of the time
arrow. To break this symmetry we transform the superfield (\ref{2.35}) to the
form (\ref{2.50}) whose explicit appearance (\ref{2.50a}) reveals breaking of
above invariance due to the Bose-Einstein condensation. Because in
macroscopical systems the time runs always forward, one needs to use the system
(\ref{2.51a}) -- (\ref{2.51d}) to describe real Fermi-Bose mixtures.

To represent the system behavior let us study first the simplest case of
steady-state homogeneous Fermi-Bose mixture. It is described by equations
(\ref{2.51a}), (\ref{2.51b}) and (\ref{2.52}) which are simplified to the
forms:
\begin{equation}
\varphi+\eta\rho={\partial F\over
\partial\eta}, \label{2.53a}
\end{equation}
\begin{equation} \label{2.53b}
{\partial^2 F\over
\partial\eta^2}\left(\varphi+\eta\rho\right)={\partial^3 F\over \partial\eta^3}
\left(1-\eta^2\right)\rho,
\end{equation}
\begin{equation}
\left(1-\eta^2\right)\rho=\rho_c.
 \label{2.53c}
\end{equation}
To reveal analytically effect of external conditions we introduce the Landau
free energy
\begin{equation}
F:=-\frac{\varepsilon}{2}\eta^2+\frac{1}{4}\eta^4,\quad
\varepsilon\equiv\frac{T_{c0}-T}{T_{c0}}
 \label{2.54}
\end{equation}
with parameter $\varepsilon$ determining a moving off a characteristic
temperature $T_{c0}$. Then, solution of Eqs. (\ref{2.53a}) -- (\ref{2.53c})
arrives at the stationary order parameter
\begin{equation}
\eta^2_0=\frac{2}{3}\varepsilon-\sqrt{\frac{1}{9}\varepsilon^2+2\rho_c}.
 \label{2.55}
\end{equation}
It takes physically meaningful magnitude $\eta_0\ne 0$ when the parameter
$\varepsilon$ is more than the critical value
\begin{equation}
\varepsilon_c=\sqrt{6\rho_c}
 \label{2.56}
\end{equation}
fixed by a critical density of fermions $\rho_c$ (together with the
characteristic temperature $T_{c0}$, above density $\rho_c$ represents
phenomenological parameter of the theory developed).

The expression (\ref{2.55}) takes the familiar square-root form
\begin{equation}
\eta_0\simeq\sqrt{\frac{T_c-T}{2T_{c0}}} \label{2.57}
\end{equation}
in the $T_c-T\ll T_{c0}$ vicinity of the critical temperature
\begin{equation}
T_c\equiv\left(1-\sqrt{6\rho_c}\right)T_{c0}. \label{2.57a}
\end{equation}
Here, Eqs. (\ref{2.53a}) -- (\ref{2.53c}) give the stationary amplitude of
fluctuations
\begin{equation}
\varphi_0\simeq-\left(\rho_c+\sqrt{6\rho_c}\right) \sqrt{\frac{T_c-T}{2T_{c0}}}
 \label{2.58}
\end{equation}
and related density of fermions
\begin{equation}
\rho_0\simeq\rho_c\left(1+\frac{T_c-T}{2T_{c0}}\right).
 \label{2.59}
\end{equation}
According Eqs. (\ref{2.57}) and (\ref{2.59}), with the temperature decrease
near the critical temperature (\ref{2.57a}) the densities of both fermions and
bosons grow linearly, while the fluctuation amplitude (\ref{2.58}) takes
negative magnitudes varying in square root manner.

Let us consider finally inhomogeneous steady-state system that is described by
the relation (\ref{2.53c}) together with the equations
\begin{equation}
\nabla^2\eta=-\left(\varepsilon-\eta^2\right)\eta-\left(\varphi+\eta\rho\right),
\label{2.60a}
\end{equation}
\begin{equation} \label{2.60b}
\nabla^2\varphi=-\left(\varepsilon-3\eta^2\right)\left(\varphi+\eta\rho\right)
-6\left(1-\eta^2\right)\eta\rho,
 \end{equation}
following from Eqs. (\ref{2.51a}) and (\ref{2.51b}) where the Landau free
energy (\ref{2.54}) is used. Linearization of Eqs. (\ref{2.60a}) and
(\ref{2.60b}) over $\eta$ and $\varphi$ shows the $\varepsilon$ increase
arrives initially (at $\varepsilon=0$) at loss of the homogeneity in space
distribution of the fluctuation amplitude and then (at
$\varepsilon=\varepsilon_c$) the system becomes unstable with respect to the
Bose-Einstein condensation. What about the fermion distribution, it is supposed
to be homogeneous due to the equilibrium condition ${\bf j}=0$ in the
continuity equation (\ref{2.52}) (consideration of more general steady-state
condition ${\bf j}={\rm const}\ne 0$ is out of the scope of our study). From
physical point of view, above means that with cooling of the Bose-Fermi mixture
characterized by a homogeneous distribution of fermions strong inhomogeneous
fluctuations appear spontaneously at a characteristic temperature $T_{c0}$,
while following temperature decrease arrives at the Bose-Einstein condensation
in the critical point $T_c$ determined by Eq.(\ref{2.57a}).

\section*{Acknowledgement}

We are grateful to Dr. Valerii A. Brazhnyi for bringing to our attention the
problem studied and discussions.

\section*{Appendix}
 \def\theequation{{A}.\arabic{equation}}
 \setcounter{equation}{0}

With accounting definitions (\ref{2.35}) and (\ref{2.37}), one obtains l.h.s.
of Eq.(\ref{2.38}):
\begin{equation}
{1\over 2} [\overline {\cal D},{\cal
D}]\Phi=\left(\varphi-\dot{\eta}\right)-\overline{\theta}\dot{\Psi}
+\dot{\overline{\Psi}}\theta +\overline{\theta}\theta\dot{\varphi}.
 \label{2.38a}
\end{equation}
Being supersymmetry variational derivative, r.h.s. of this equation is written
as the generalization of the expression (\ref{2.4}):
\begin{equation} \label{2.38b}
\begin{split}
{\delta\mathcal{F}\over \delta \Phi}&=F'\left(\eta+\overline\theta\Psi
+\overline\Psi\theta + \overline\theta\theta\varphi\right)\\
&-\left[\nabla^2\eta+\overline\theta\left(\nabla^2\Psi\right)
+\left(\nabla^2\overline{\Psi}\right)\theta +
\overline\theta\theta\left(\nabla^2\varphi\right)\right]
\end{split}
\end{equation}
where the prime denotes the differentiation over related argument. According to
the first rule (\ref{2.12b}), expansion in powers of the addition
$\overline\theta\Psi +\overline\Psi\theta + \overline\theta\theta\varphi$ gives
\begin{equation} \label{2.38c}
\begin{split}
F'\left(\Phi\right)&=F'\left[\eta+\left(\overline\theta\Psi
+\overline\Psi\theta
+\overline\theta\theta\varphi\right)\right]\\&=F^{'}(\eta)+F^{''}(\eta)\left(\overline\theta\Psi
+\overline\Psi\theta\right)\\&+\left[F^{''}(\eta)\varphi-
F^{'''}(\eta)\rho\right]\overline\theta\theta.
\end{split}
\end{equation}
Comparison of multipliers standing before $1$, $\overline\theta$, $\theta$ and
$\overline\theta\theta$ arrives at the system of equations (\ref{2.39a}) --
(\ref{2.39d}).

In more complicated case of the transformed superfield (\ref{2.50a}), the
expressions (\ref{2.38a}) and (\ref{2.38c}) take the forms:
\begin{equation} \label{2.38a1}
\begin{split}
&{1\over 2} [\overline {\cal D},{\cal
D}]\widetilde\Phi=\left[\left(\varphi-\dot{\eta}\right)+\eta\rho\right]\\
&-\left[\left(1-\eta\right)\frac{\dot{\Psi}}{\Psi}-\dot{\eta}\right]
\overline{\theta}\Psi+\left[\left(1+\eta\right)\frac{\dot{\overline\Psi}}{\overline\Psi}
+\dot{\eta}\right]\overline{\Psi}\theta\\ &+ \frac{\partial}{\partial
t}\left(\varphi+\eta\rho\right)\overline{\theta}\theta,
\end{split}
\end{equation}
\begin{equation} \label{2.38c1}
\begin{split}
&F'\left(\widetilde\Phi\right)=F^{'}(\eta)\\&+F^{''}(\eta)\left(1-\eta\right)\overline\theta\Psi
+F^{''}(\eta)\left(1+\eta\right)\overline\Psi\theta\\&+
\left[F^{''}(\eta)\left(\varphi+\eta\rho\right)-
F^{'''}(\eta)\left(1-\eta^2\right)\rho\right]\overline\theta\theta.
\end{split}
\end{equation}
Comparison of related supersymmetry terms derives to the system (\ref{2.51a})
-- (\ref{2.51d}).

\end{document}